\documentclass[aps,reprint,twocolumn,showpacs,superscriptaddress]{revtex4-1}
\usepackage[colorlinks,linkcolor=blue,anchorcolor=blue,citecolor=blue,filecolor=blue,menucolor=blue,runcolor=blue,urlcolor=blue,frenchlinks=blue]{hyperref}
\usepackage{graphicx}
\usepackage{subfigure}
\usepackage{amsmath}
\usepackage{amssymb}
\usepackage{dcolumn}
\usepackage{mathrsfs}
\usepackage{bm,physics}
\usepackage{color,amsfonts,amsthm,mathtools}
\usepackage{caption}
\usepackage{subfigure}

\makeatletter

\newcommand{\Rmnum}[1]{\expandafter\@slowromancap\romannumeral #1@}
\makeatother

\begin{document}
\title{Tensor monopoles and negative magnetoresistance effect in optical lattices}

\author{Hai-Tao Ding}
\affiliation{National Laboratory of Solid State Microstructures and School of Physics, Nanjing University, Nanjing 210093, China}

\author{Yan-Qing Zhu}
\email{dg1722056@smail.nju.edu.cn}
\affiliation{National Laboratory of Solid State Microstructures and School of Physics, Nanjing University, Nanjing 210093, China}

\author{Zhi Li}\affiliation{Guangdong Provincial Key Laboratory of Quantum Engineering and Quantum Materials, School of Physics and Telecommunication Engineering, South China Normal University, Guangzhou 510006, China}

\affiliation{Guangdong-Hong Kong Joint Laboratory of Quantum Matter, South China Normal University, Guangzhou 510006, China}

\author{Lubing Shao}
\email{lbshao@nju.edu.cn}
\affiliation{National Laboratory of Solid State Microstructures and School of Physics, Nanjing University, Nanjing 210093, China}

\date{\today}

\begin{abstract}
  We propose that a kind of four-dimensional (4D) Hamiltonians, which host tensor monopoles related to quantum metric tensor in even dimensions, can be simulated by ultracold atoms in the optical lattices. The topological properties and bulk-boundary correspondence of tensor monopoles are investigated in detail. By fixing the momentum along one of the dimensions, it can be reduced to an effective three-dimensional model manifesting with a nontrivial chiral insulator phase. Using the semiclassical Boltzmann equation, we calculate the longitudinal resistance
against the magnetic field $B$ and find the negative relative magnetoresistance effect of approximately $ -B^{2} $ dependence when a hyperplane cuts through the tensor monopoles in the parameter space. We also propose an experimental scheme to realize this 4D Hamiltonian by introducing an  external cyclical parameter in a 3D optical lattice. Moreover, we show that the quantum metric tensor and Berry curvature can be detected by applying an external drive in the optical lattices.
\end{abstract}

\maketitle

\section{Introduction}

In 1931, Dirac introduced the concept of monopoles to explain the quantization of electron charge~\cite {MDirac}. Since then, the development of gauge theory has shown that monopoles emerge in a natural way in all theories of grand unification. However, the existence of the monopole as an elementary particle has not been confirmed by any experiment till today. Monopoles in momentum space have attracted extensive studies in condensed matter physics and artificial quantum systems, such as, Dirac monopoles in Weyl semimetals. The celebrated Nielsen-Ninomiya theorem states that Weyl points in the first Brilluoin zone must emerge and annihilate in pairs with opposite chirality, which provides a mechanism of anomaly cancellation in the field theorem framework~\cite{MNinomiya}.
Moreover, negative magneto-resistance (MR) effect, for which the longitudinal conductivity increases along with the increasing magnetic field, has been reported in several experiments and can be interpreted as a result of the suppression of backscattering due to the chirality of the monopoles in Weyl semimetals~\cite {Shen,Sun,longitudinal,Son,DWZhang2016,Positive,pyrochlore}.
Besides those monopoles in odd dimensions, recent research shows that another kind of monopoles can emerge in even dimensions, named ``tensor monopoles", which are Abelian monopoles associated with the tensor (Kalb-Ramond) gauge field~\cite {GaPalumbo}. The topological charge of the tensor monopole is related to the so-called quantum metric tensor which measures the distance of two nearby states in the parameter space. Recently, by using controllable quantum systems, several experiments have been reported to directly measure the quantum metric tensor, which characterizes the geometry and topology of underlying quantum states in parameter space~\cite{Zhang,and,yu2020experimental,gianfrate2020measurement}. { More recently, two experiments for realizing the tensor monopoles in 4D parameter space have been reported in superconducting circuit system\cite{XTan2020} and the nitrogen-vacancy (NV) center in diamond\cite{MChen2020}, respectively.}

The technology of ultracold atoms provides an excellent platform to study different topological systems of condensed matter and high-energy physics, because of its perfect cleanness and high controllability~\cite{zhang2018topological}. Recently, 4D quantum Hall effect has also be experimentally simulated by ultracold atoms, which opens up the research of high-dimensional physics in realistic systems~\cite{Carusotto,Schweizer}. It shows that {an extra dimension} can be engineered by a set of internal atomic levels as {a synthetic lattice dimension}~\cite{Boada,Phys}.
{The 4D Hamiltonian can also be realized in a 3D optical lattice with a cyclical parameter  which playing the role of the pseudo-momentum of the fourth dimension~\cite{SLZhu2013,Sarma,Nature,FMei2012,YLChen2020,DWZhang2020}.}  In order to measure the Berry phase of topological systems in cold atoms, many experimental approaches have been proposed and conducted, including state tomography~
\cite{Rem,Duca}, interferometry~\cite{79}, and atomic transport~\cite{80}. Recent development on how to measure the quantum metric tensor and Berry curvature by shaking the optical lattice has promoted the research of tensor monopoles with cold atoms~\cite{and,Tran}.

In this paper, we propose two minimal Hamiltonians in 4D which host tensor monopoles~\cite {GaPalumbo,Teitelboim,Orland,Kalb,GrPalumbo}, and then study their topological properties. Tensor monopoles in these two systems can be considered as one conductance band, one valence band and one flat band touching at the common points and the topological properties of the tensor monopoles can be controlled by a tunable parameter. After fixing the momentum of the fourth dimension in the parameter space, we obtain a 3D model. Rich phase diagrams can be derived from this 3D model, including trivial phase and chiral insulator phase. In addition, we calculate the MR with the semiclassical Boltzmann equation. When a hyperplane cuts through the tensor monopoles, the relative MR approximately proportional to $-B^2$ signifies the negative MR effect, where $ B $ is the magnetic field.
We also propose an experimentally feasible scheme to implement our model with three-component ultracold atoms in a 3D optical lattice with { an external parameter varying from $-\pi$ to $\pi$ which can be treated as the fourth dimension}. Moreover, we provide the experimental method to measure quantum metric tensor and Berry curvature.

The paper is organized as follows. In section \uppercase\expandafter{\romannumeral2}, we introduce the models for realizing tensor monopoles and review the definition of the quantum metric tensor with  topological charge. In section \uppercase\expandafter{\romannumeral3}, we give the tight-binding Hamiltonians and investigate the bulk-boundary correspondence. In section \uppercase\expandafter{\romannumeral4}, MR effect is calculated for our tensor monopole models, using the semiclassical Boltzmann equation. In section \uppercase\expandafter{\romannumeral5}, an experimental scheme to realize the 4D model is established with a proposal of measuring the quantum metric tensor and Berry curvature. Finally, a brief conclusion is provided in section \uppercase\expandafter{\romannumeral6}.

\section{Tensor monopoles in 4D flat space}
Following Ref.~\cite {MaJGilbert}, the tensor monopole in momentum space can be hosted by a generalization of 4D multi-Weyl Hamiltonian as
\begin{equation}
\begin{split}
\label{E1}
H_n=\frac{1}{2}(k_{-}^n\lambda_{+}+k_{+}^n\lambda_{-})+\alpha_z k_z\lambda_{6}+\alpha_w k_w\lambda_{7}^{*},
\end{split}
\end{equation}
where $\textbf{k}=(k_x,k_y,k_z,k_w)$ is the 4D momentum. The $\lambda_i(i=1,2,6,7)$ are Gell-Mann  matrices~\cite {MaGell-Mann}, which are representations of the infinitesimal generators of $\text{SU(3)}$. Here we set $k_{\pm}= k_{x}\pm \mathrm{i}k_{y}$ and $\lambda_{\pm}= \alpha_x \lambda_{1}\pm i\alpha_y \lambda_{2}$, $\alpha_j=\pm1(j=x,y,z,w)$. The Hamiltonian breaks time reversal symmetry, but keeps chiral symmetry due to the anticommutative relation $\{U,H_n\}=0$  with $U=\text{diag}(1,-1,1)$. Energy spectra are obtained as
\begin{equation}
E_{\pm}=\pm\sqrt{\left(k_{x}^{2}+k_{y}^{2}\right)^{n}+k_{z}^{2}+k_{w}^{2}},\qquad E_{0}=0.
\label{spect}
\end{equation}
And the related eigenstates are denoted by $\left|u_{\pm}\right\rangle$, $\left|u_{0}\right\rangle$.
The tensor monopole exists at $\textbf{k}=(0,0,0,0)$, where the three bands touch commonly.

Recall that the Dirac monopoles and non-Abelian Yang monopoles are defined in 3D and 5D parameter spaces, respectively. They are all described by vector Berry connections, i.e., vector gauge field. But for tensor monopoles defined in 4D parameter space, they are captured by tensor Berry connection. The associated gauge field is an Abelian antisymmetric tensor field $B_{\mu \nu}$ called the Kalb-Ramond field, which is defined as
\begin{equation}
\begin{split}
&B_{\mu \nu}=\phi F_{\mu \nu}, \\
&\phi=-\frac{i}{2} \log \prod_{\aleph=1}^{3} u_{-}^{\aleph},
\end{split}
\end{equation}
Here $\phi$ is a pseudo-scalar gauge field, $F_{\mu \nu}=\partial_{\mu} A_{\nu}-\partial_{\nu} A_{\mu}$ is Berry curvature($\partial_{\mu} \equiv \partial_{k_{\mu}}$), and the associated Berry connection $A_{\mu}=\left\langle u_{-}\left|i \partial_{\mu}\right| u_{-}\right\rangle$, $u_{-}^{\aleph}$ denotes the components of the lowest band $\left|u_{-}\right\rangle$~\cite{YQZhu2020}.  Related 3-form curvature is $\mathcal{H}=d B$, whose components are given by
\begin{equation}\mathcal{H}_{\mu \nu \lambda}=\partial_{\mu} B_{\nu \lambda}+\partial_{\nu} B_{\lambda \mu}+\partial_{\lambda} B_{\mu \nu}.\end{equation}
It is gauge invariant and antisymmetric. For Hamiltonian in Eq.~\eqref{E1}, the corresponding 3-form curvature is
\begin{equation}\begin{aligned}
\mathcal{H}_{\mu \nu \lambda}=\text{sgn}(\alpha_x \alpha_y \alpha_z \alpha_w)\epsilon_{\mu \nu \lambda \gamma}\frac{nk_{\gamma}}{\left(k_{x}^{2}+k_{y}^{2}+k_{z}^{2}+k_{w}^{2}\right)^{2}}.
\end{aligned}\end{equation}

 A topological charge associated with this curvature $\mathcal{H}_{\mu \nu \lambda}$ can be defined by surrounding the tensor monopole with a sphere $S^3$,
\begin{equation}
\label{E6}
Q_{n}=\frac{1}{2 \pi^{2}} \int_{S^{3}} d k^{\mu} \wedge d k^{\nu} \wedge d k^{\lambda} \mathcal{H}_{\mu \nu \lambda}.\end{equation}
This is a topological invariant known as the Dixmier-Douady($DD$) invariant, which is related to the (first) Dixmier-Douady class of U(1) "bundle gerbes"~\cite{Mathai,Murray,Geometry,Handbook}.

{Directly measuring the tensor Berry curvature in the experiment actually is difficult. Fortunately,  we can find a direct relation between the components of 3-form curvature and the quantum metric(or Fubini-Study metric)
\cite{GaPalumbo}},
\begin{equation}\label{TenBerry}
\mathcal{H}_{\mu \nu \lambda}=\text{sgn}(\alpha_x \alpha_y \alpha_z \alpha_w)\epsilon_{\mu \nu \lambda}(4 \sqrt{\operatorname{det} g_{\bar{\mu} \bar{\nu}}}),
\end{equation}
where $g_{\bar{\mu} \bar{\nu}}$ is the $3\times3$ quantum-metric tensor defined in the proper 3D subspace. Physically, if the Hamiltonian of a system is parametrized as $ H\equiv H(\vec{\lambda}) $, quantum metric tensor measures the (infinitesimal) distance between two nearby quantum states, $ d s^{2}=1-\left|\left\langle \psi_{\lambda} | \psi_{\lambda+d \lambda}\right\rangle\right|^{2} $, in $\vec{\lambda}$ space \cite {Provost} as
\begin{equation}
d s^{2}=\sum_{\mu \nu} g_{\mu \nu} d \lambda_{\mu} d \lambda_{\nu},
\end{equation}
in which the metric tensor can be explicitly written as
\begin{equation}
\label{E2}
g_{\mu \nu }=\text{Re}\left(\braket{\frac{\partial \psi }{\partial \lambda_{\mu} }}{\frac{\partial \psi }{\partial \lambda_{\nu} }}-\braket{\frac{\partial \psi }{\partial \lambda_{\mu} }}{\psi}\braket{\psi}{\frac{\partial \psi }{\partial \lambda_{\nu} }}\right).
\end{equation}
Obviously, the metric is positive and satisfies $g_{\mu \nu }=g_{\nu \mu}$.
{As a consequence, Eq. \eqref{TenBerry} provides a feasible method to detect the 3-form curvature through quantum metric tensor measurement, which will be discussed in section \uppercase\expandafter{\romannumeral5}.}

{In other words, one can calculate $DD$ invariant through the quantum metric tensor. To be more clear,  we show some derivations of $DD$ invariant by calculating the quantum metric below.} For simplify,  we can parameterize the momentum space $\textbf{k}=(k_x,k_y,k_z,k_w)$ in Eq.~\eqref{E1} with the hyperspherical coordinates $(k,\theta_{1},\theta_{2},\varphi)$ as
\begin{equation}
\begin{split}
k_{x}&=(k \sin\theta_{1}\sin\theta_{2})^{\frac{1}{n}}\cos\varphi,\\
k_{y}&=(k \sin\theta_{1}\sin\theta_{2})^{\frac{1}{n}}\sin\varphi,\\
k_{z}&=k \sin\theta_{1}\cos\theta_{2},\\
k_{w}&=k \cos\theta_{1},
\end{split}
\label{hs}
\end{equation}
where $k=\sqrt{\left(k_{x}^{2}+k_{y}^{2}\right)^{n}+k_{z}^{2}+k_{w}^{2}}$ is the radius of the 3-hypersphere encircling the monopole in momentum space. If the lowest energy band $\varepsilon_{\mathbf{k}}=-k$ is filled, the topological charge of the tensor monopole can be defined as
\begin{equation}\begin{aligned}
Q_{n} =\frac{1}{2 \pi^{2}} \int_{0}^{\pi} d \theta_{1} \int_{0}^{\pi} d \theta_{2} \int_{0}^{2 \pi} d \varphi \mathcal{H}_{\theta _1 \theta _2 \varphi },
\end{aligned}\end{equation}
where $ \mathcal{H}_{\theta _1 \theta _2 \varphi }=4\text{sgn}(\alpha_x \alpha_y \alpha_z \alpha_w) \sqrt{\det g} $. By considering the hypersphere surrounding the monopole, we derive $ \det g=\frac{1}{16}n^2 \sin^4 \theta _1 \sin^2 \theta _2 $ and the topological charge is obtained as
\begin{equation}\label{EE5}
Q_{n }=n\text{sgn}(\alpha_x \alpha_y \alpha_z \alpha_w).
\end{equation}

 We consider two special cases of the Hamiltonians in Eq.~\eqref{E1} as
\begin{equation}\label{E5}
H_1=k_x\lambda_1+k_y\lambda_2+k_z\lambda_6+k_w\lambda_{7}^{*},
\end{equation}
for $ n=1 $, and
\begin{equation}\label{E6}
H_2=\left(k_{x}^{ 2}-k_{y}^{ 2}\right) \lambda_{1}+2 k_{x} k_{y}\lambda_{2}+ k_{z} \lambda_{6}+ k_{w} \lambda_{7}^{*},
\end{equation}
for $ n=2 $.
In both cases, the three energy bands all crossing at the $\textbf{k}=(0,0,0,0)$, which hosts tensor monopoles with the topological charges being $Q_1=1$ and $Q_2=2$, respectively.

\section{The minimal models in momentum space}
Now we construct the Hamiltonians in Eq.~\eqref{E5} and \eqref{E6} with the tight-binding models in the momentum space as
\begin{equation}
\label{tb-H}
\mathcal{H}_{n}=d_{n,x}\lambda_1+d_{n,y}\lambda_2+d_{n,z}\lambda_6+d_{n,w}\lambda_{7}^{*},
\end{equation}
with $n=1,2$.

\begin{figure}\centering
	\includegraphics[width=8.5cm]{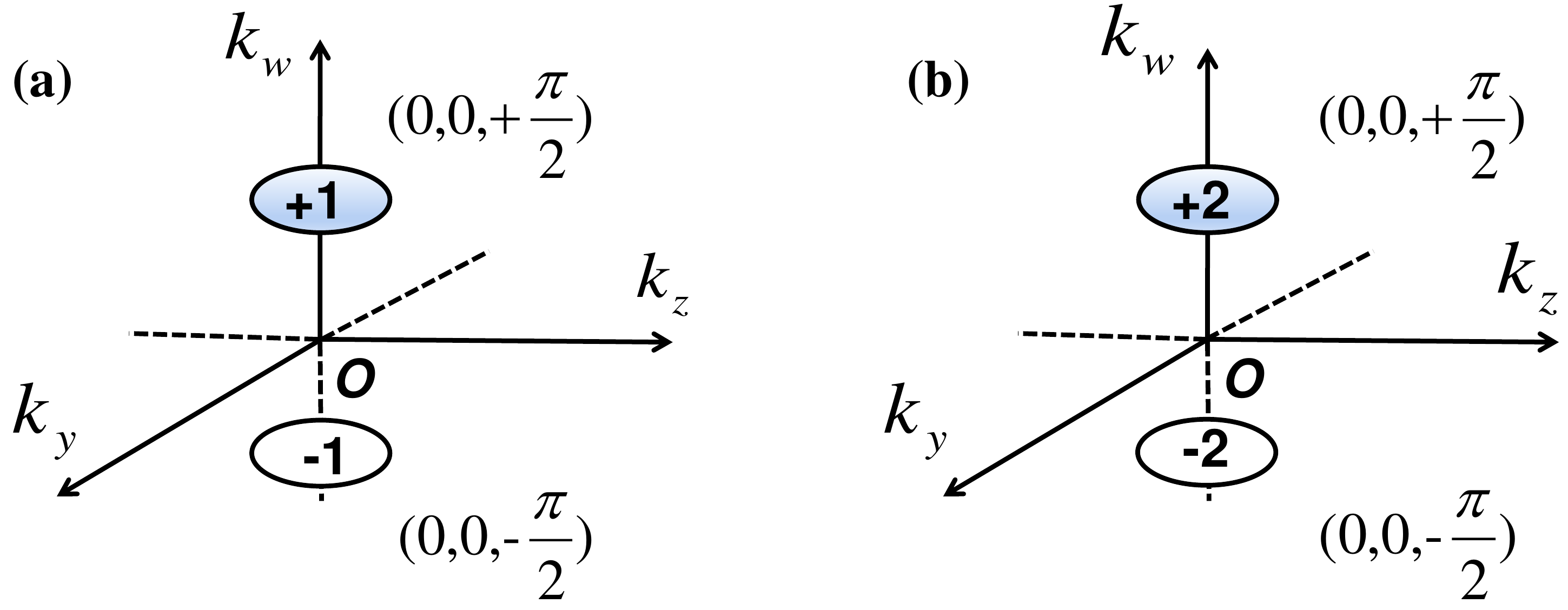}
	\caption{(a) and (b) are the monopoles of $\mathcal{H}_1$ and $\mathcal{H}_2$ for $h=3$. We plot the monopoles in $k_y-k_z-k_w$ parameter space and define $k_x=0$. The ellipses stand for monopoles and the numbers mean the topological charges of them.}
	\label{fig1}
\end{figure}

For $ n=1 $, the explicit form of $d$'s are
\begin{equation}
\begin{split}
d_{1,x}&=2t\sin k_x, \\
d_{1,y}&=2t\sin k_y, \\
d_{1,z}&=2t\sin k_z, \\
d_{1,w}&=2t(h-\cos k_x-\cos k_y-\cos k_z-\cos k_w),
\end{split}
\label{asd}
\end{equation}
where we have set the lattice constant $a=1$. Here, $t$ is hopping energy and $h$ is a tunable parameter. The corresponding spectrum is given by $\left\{0,\pm\sqrt{d_{1,x}^2+d_{1,y}^2+d_{1,z}^2+d_{1,w}^2}\right\}$. When $ h=3 $, there exist a pair of triple-degenerate Dirac-like points at $\textbf{K}_{\pm}=(0,0,0,\pm\pi/2)$, which are tensor monopoles with topological charges $ \pm1 $, as shown in Fig.~\ref{fig1}(a) with $ k_x=0 $. The $ k\cdot p $ Hamiltonian near the two nodes with $\textbf{q}=\textbf{k}-\textbf{K}_{\pm}$ yields the low-energy effective Hamiltonian
as Eq.~\eqref{E5}.

Similarly, for $ n=2 $, the $ d $'s can be written as
\begin{equation}
\begin{split}
d_{2,x}&=2t(\sin^2 k_x-\sin^2 k_y),\\
d_{2,y}&=4t\sin{k_x}\sin{k_y}, \\
d_{2,z}&=2t\sin{k_z}, \\
d_{2,w}&=2t(h-\cos{k_x}-\cos{k_y}-\cos{k_z}-\cos{k_w}),
\end{split}
\label{asc}
\end{equation}
by which the energy dispersion is obtained as $\left\{0,\pm\sqrt{d_{2,x}^{ 2}+d_{2,y}^{ 2}+d_{2,z}^{ 2}+d_{2,w}^{ 2}}\right\}$. For $h=3$, there are also a pair of tensor monopoles at $\textbf{K}_{\pm}=(0,0,0,\pm\pi/2)$ with topological charges $ \pm2 $ as shown in Fig.~\ref{fig1}(b) with $ k_x=0 $. The low-energy effective Hamiltonian near the two nodes is obtained as Eq.~\eqref{E6}.

For both cases of $ n=1 $ and $ n=2 $, the combination and division of tensor monopoles inside the first Brillouin zone (FBZ) are controlled by the parameter $h$.
For $h=0$, there are six trivial monopoles located at $(\pi,\pi,0,0)$, $(\pi,0,\pi,0)$, $(\pi,0,0,\pi)$, $(0,\pi,\pi,0)$, $(0,\pi,0,\pi)$ and $(0,0,\pi,\pi)$. Increasing $ h $, the six monopoles begin to move in FBZ. When $h=1$, the six degenerate points move to $(0,0,\pi,\pm\pi/2)$, $(0,\pi,0,\pm\pi/2)$, $(\pi,0,0,\pm\pi/2)$,  including three of them with positive topological charge and three others with negative topological charge as a result of the generalized Nielsen-Ninomiya theorem~\cite{MNinomiya}. When continuously increasing $ h $ to $h=2$, there are four monopoles left at $(\pi,0,0,0)$, $(0,\pi,0,0)$, $(0,0,\pi,0)$, $(0,0,0,\pi)$, which are also trivial. For $h=3$, only two monopoles are left at $(0,0,0,\pm\pi/2)$. For $h=4$, the two monopoles move toward $(0,0,0,0)$ and combine to open a gap. Finally, it becomes a topologically-trivial insulator for $ h>4 $.

By taking a slice of these two 4D models, i.e., fixing $k_w=0$, 3D models can be derived from the 4D systems. The topological nature of the 3D system is captured by DD invariant. This invariant is equivalent to the winding number, which characterizes 3D topological insulators in class $AIII$~\cite {Ne,Palumbo,Tang}
\begin{equation}\begin{aligned}
\label{E8}
\Gamma_{n} &=\frac{1}{12\pi ^2}\int_{BZ}d^3k \epsilon^{\alpha\beta\gamma\rho} \epsilon^{\mu\nu\tau}\frac{1}{E_{+}^4}d_{\alpha}\partial_{\mu}d_{\beta}\partial_{\nu}d_{\gamma}\partial_{\tau}d_{\rho} \\
&=Q_{n},
\end{aligned}\end{equation}
where $E_{+}(k)=\sqrt{d_{n,x}^2+d_{n,y}^2+d_{n,z}^2+d_{n,w}^2}$, and the indexes of the Levi-Civita symbol with $\alpha,\beta,\gamma,\rho$ and $\mu,\nu,\tau$ represent $\{x,y,z,w\}$ and $\{k_x,k_y,k_z\}$, respectively.

\begin{figure}
\centering
 \includegraphics[width=8.7cm]{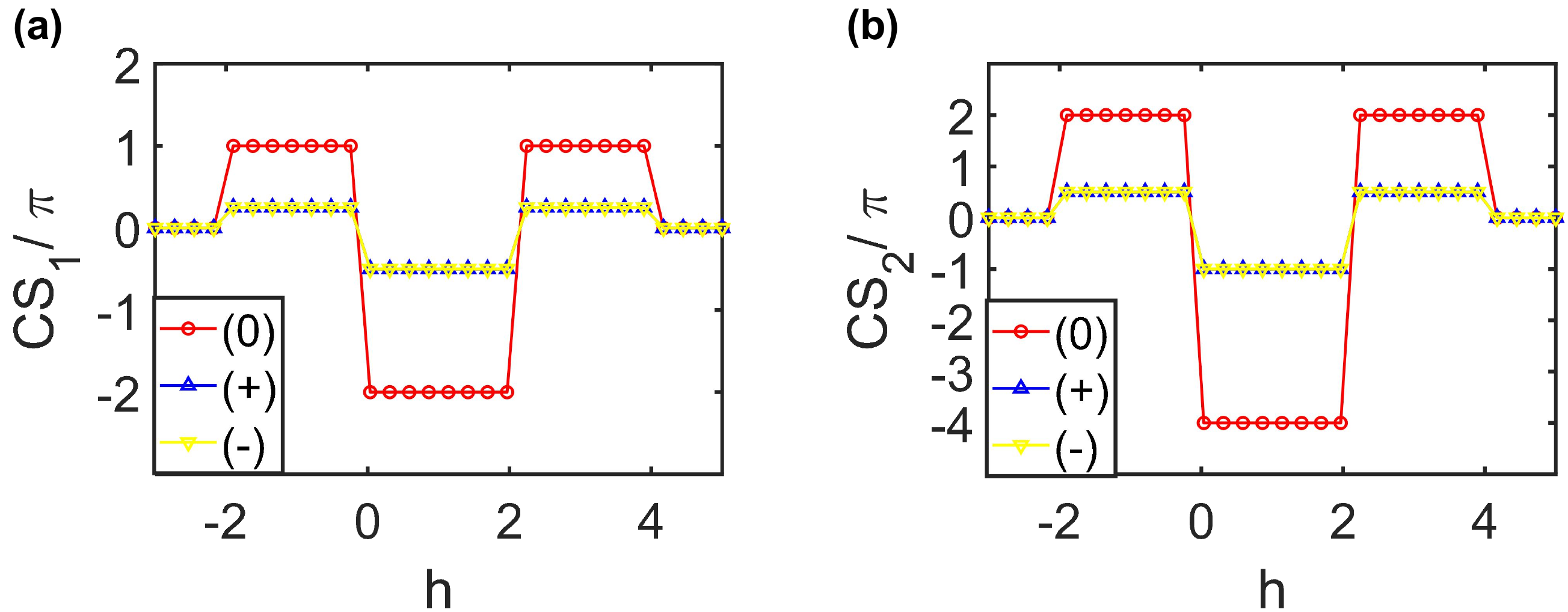}
 \caption{(a) and (b) stand for the Chern-Simons term of $\mathcal{H}_1$ and $\mathcal{H}_2$ respectively. $'(+)'$ stands for the positive energy band, $'(0)'$ and $'(-)'$ represent the flat and negative energy band respectively. Here we set $t=1$. }
 \label{fig2}
\end{figure}

Another equivalent way to characterize the topology of the 3D models is the Chern-Simons invariant (CSI), which takes the form as
\begin{equation}
\mathrm{CS}=\frac{1}{4 \pi} \int_{\mathrm{BZ}} d \mathbf{k} \epsilon^{\mu \nu \tau} A_{\mu}(\mathbf{k}) \partial_{\nu} A_{\tau}(\mathbf{k}),
\end{equation}
where $A_{\mu}(\mathbf{k})=\left\langle u(\mathbf{k})\left|i \partial_{\mu}\right| u(\mathbf{k})\right\rangle(\mu=x, y, z)$~\cite{Ne,Aa}. We plot CSI against $h$ in Fig.~\ref{fig2}(a) with $ n=1 $ and \ref{fig2}(b) with $ n=2 $ for the three energy bands at $ k_w=0 $. The relation of the value of it between different bands is
\begin{equation}
CS_n (+)=CS_n (-)=\frac {1}{4}CS_n (0).
\end{equation}
As indicated in Fig.~\ref{fig2}, the topological phase transitions occur at $h=0,\pm 2,4$, when the three bands touch at the hyperplane of $ k_w=0 $.
 For $h\in{(-2,4)}$, the $CS_n$ is nonzero, it is topologically nontrivial phases here.
 A detailed calculation shows that the relation between winding number and the Chern-Simons term is
\begin{equation}
\frac {\pi}{4}\Gamma_n=CS_n(-).
\end{equation}

\begin{figure}
	\centering
	\includegraphics[width=8.5cm]{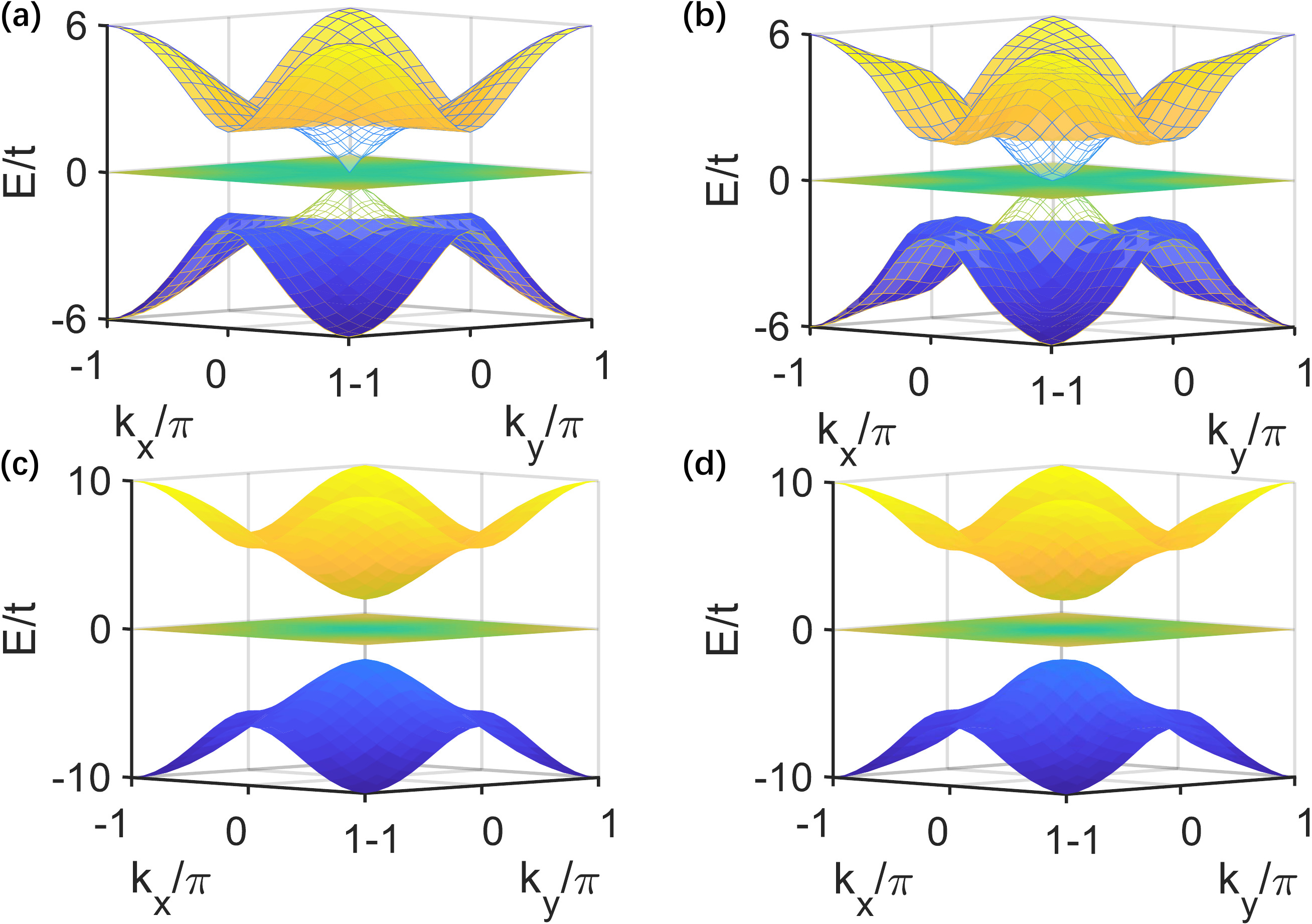}
	\caption{The bulk state(surface plot) and surface state(mesh plot), for $k_w=0$. (a) is the case of $\mathcal{H}_1, h=3$. (b) is $\mathcal{H}_2, h=3$. They are chiral topological insulator phase, the surface state connects three gapped bands. (c) and (d) are the bulk state of $\mathcal{H}_1, h=5$ and $\mathcal{H}_2, h=5$. They are trivial insulators.}
	\label{fig3}
\end{figure}

When $ h=5 $, according to Fig.~\ref{fig2}, the 3D systems is trivial and no surface state exists, which is confirmed by the numerical calculation shown in Fig.~\ref{fig3} (c) and (d).

\begin{figure}
	\centering
	\includegraphics[width=8.5cm]{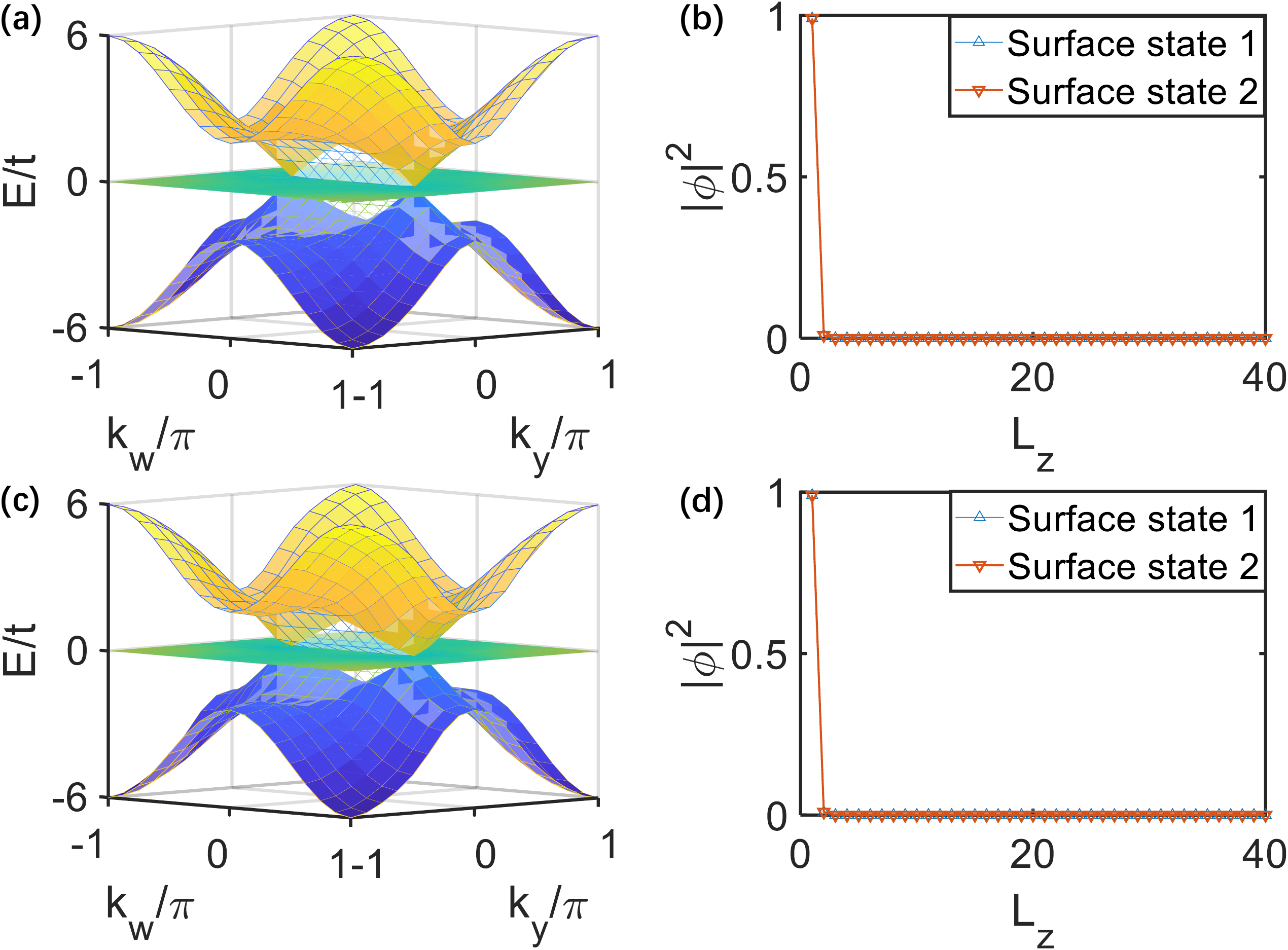}
	\caption{(a) and (c) are the bulk state(surface plot) and surface state (mesh plot) of $\mathcal{H}_1$ and $\mathcal{H}_2$ respectively.  (b) and (d) are the density distribution of the wave function of surface state for $\mathcal{H}_1$ and $\mathcal{H}_2$ respectively. Here $k_x=0$, $k_y=k_w=0.1\pi$.}
    \label{fig4}
\end{figure}
The energy spectrum and surface states with the open boundary along $ \hat{z} $ direction are shown in Fig.~\ref{fig3}. As discussed before, for $h=3$, two tensor monopoles are located at $ (0,0,0,\pm\pi/2) $ and the spectrum is gapped and topologically nontrivial for $ k_{w}\in (-\pi/2,\pi/2)$. They are chiral insulator phases. Therefore, there are surface states of Dirac cones for those sliced 3D systems until the slicing hyperplane hits the tensor monopoles. Namely, surface Dirac cone survives when $ -\pi/2<k_w<\pi/2 $. The spectra with $ k_w=0 $ are shown in Fig.~\ref{fig3} (a) and (b) for $ n=1 $ and $ n=2 $, respectively. When we take the slicing hyperplane perpendicular to $ \hat{x} $ axis, the sets of those Dirac points constitute the Fermi arcs connecting two tensor monopoles for $ n=1 $ and $ n=2 $, which are shown in Fig.~\ref{fig4} (a) and (c), respectively. Fig.~\ref{fig4} (b) and (d) shows the density distribution of surface states. By using the method in Ref~\cite{Ne}, for $k_w \in (-\pi/2,\pi/2)$, the low-energy spectra of surface states around $(k_x,k_y,k_z)=(0,0,0)$ are $ \pm v\sqrt{k_x^2+k_y^2} $ and $ \pm v(k_x^2+k_y^2) $ for $ n=1 $ and $ n=2 $, respectively. Here $v=2t$ is the effective Fermi velocity. Detailed derivation of these spectra can be found in the Appendix.

\section{Transport property}	
Negative magnetoresistance effect has already been extensively discussed in topological semimetals.  This fantastic transport phenomena is widely believed to be caused by chiral anomaly,  which is the violation of the conservation of chiral current~\cite{longitudinal}. In some topological insulators, there also emerges the same effect, although the chiral anomaly is not well defined in these systems~\cite{Dimitrieq,Du}. By using semiclassical equation, we can calculated MR.

In semiclassical limit, the electronic transport can be described by the equations of motion 
\begin{equation}
\begin{split}
\dot{\mathbf{r}}&=\frac{1}{\hbar} \nabla_{\mathbf{k}} \tilde{\varepsilon}_{\mathbf{k}}-\dot{\mathbf{k}} \times  \mathbf{\Omega}_{\mathbf{k}}, \\
\dot{\mathbf{k}}&=-\frac{e}{\hbar}(\mathbf{E}+\dot{\mathbf{r}} \times \mathbf{B}), \\
\tilde{\varepsilon}_{\mathbf{k}}&=\varepsilon_{\mathbf{k}}-\mathbf{M} \cdot \mathbf{B},  \\
\mathbf{M}&=-\frac{e}{2h}\operatorname{Im}\bra{\frac{\partial u}{\partial \mathbf{k}}} \left(\mathcal{E}_{0}-\hat{H}_{0}(\mathbf{k})\right)\ket{\frac{\partial u}{\partial \mathbf{k}}},
\end{split}
\end{equation}
which describe the dynamics of the wave packet~\cite{Xiao,Shindou,Sundaram}. Here, $ \mathbf{r} $ is the position of the wave packet in real space, and $ \mathbf{k} $ corresponds to the wave vector. $\mathbf{\Omega}_{\mathbf{k}}$ is Berry curvature, $\varepsilon_\textbf{k}$ is the energy dispersion of the valence band, and $\mathbf{M}$ is orbital magnetic moment of the wave packet which is analogous to the magnetic moment of a electron motions around the nucleus~\cite {Valley}.

Using the semiclassical Boltzmann equation, the longitudinal conductivity can be calculated by
\begin{equation}
\label{E23}
\begin{aligned}
\sigma^{\mu \mu} &=\int \frac{d^{3} \mathbf{k}}{(2 \pi)^{3}} \frac{e^{2} \tau}{D_{\mathbf{k}}}\left(\tilde{v}_{\mathbf{k}}^{\mu}+\frac{e}{\hbar} B^{\mu} \tilde{v}_{\mathbf{k}}^{\nu} \Omega_{\mathbf{k}}^{\nu}\right)^{2}\left(-\frac{\partial \tilde{f}_{0}}{\partial \tilde{\varepsilon}}\right) \\
D_{\mathbf{k}} &=1+\frac{e}{\hbar} \mathbf{B} \cdot \mathbf{\Omega}_{\mathbf{k}}
\end{aligned}
\end{equation}
where $\tilde{f}_0$ is the equilibrium Fermi distribution, and $\tau$ is the life time of the quasiparticle in the semiclassical limit~\cite {Du,A}. $\tilde{v}_{\mathbf{k}}^{\nu}$ and $\Omega_{\mathbf{k}}^\nu$ are the components of velocity and Berry curvature.
\begin{figure}
	\centering
	\includegraphics[width=6.5cm]{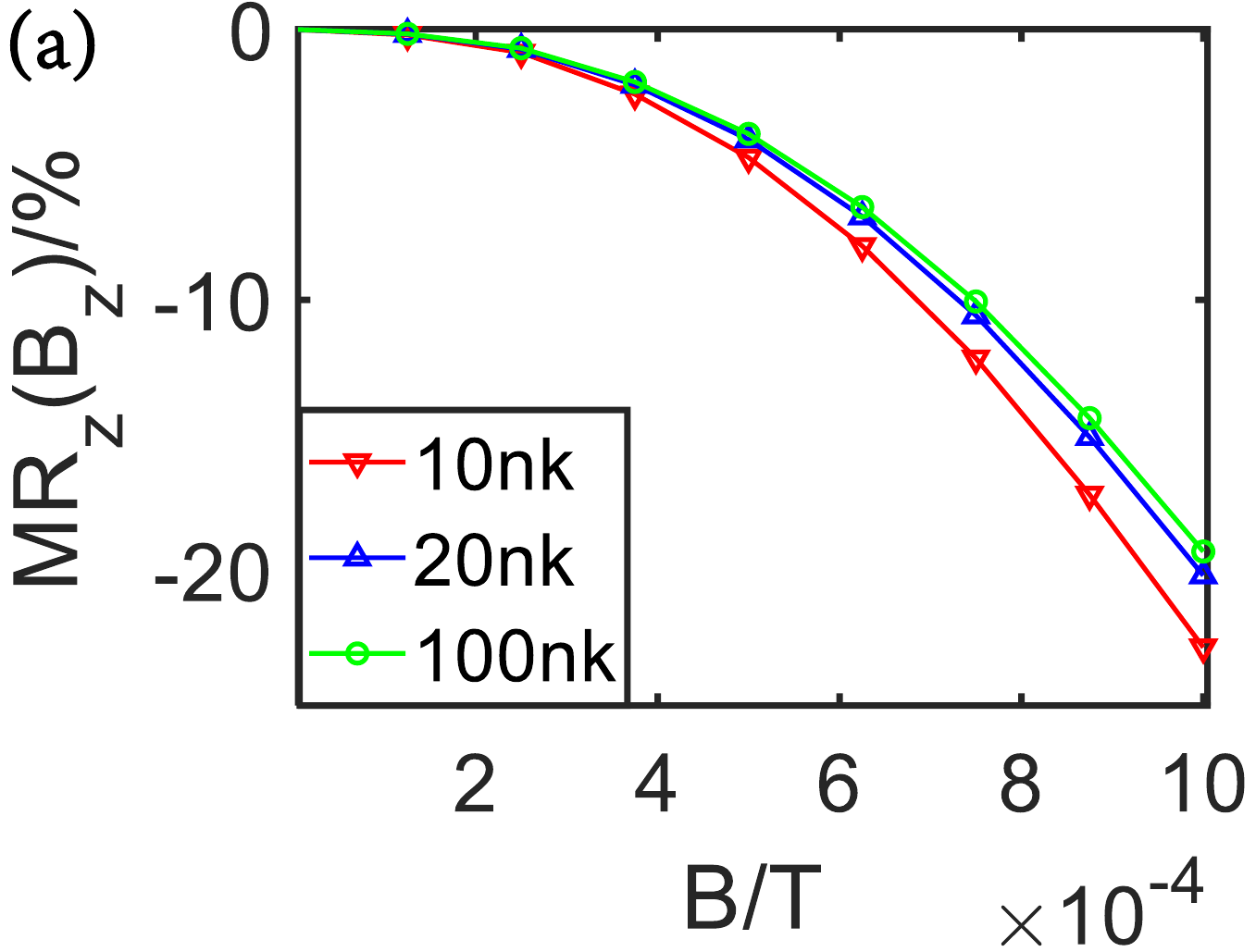}\ \ \includegraphics[width=6cm]{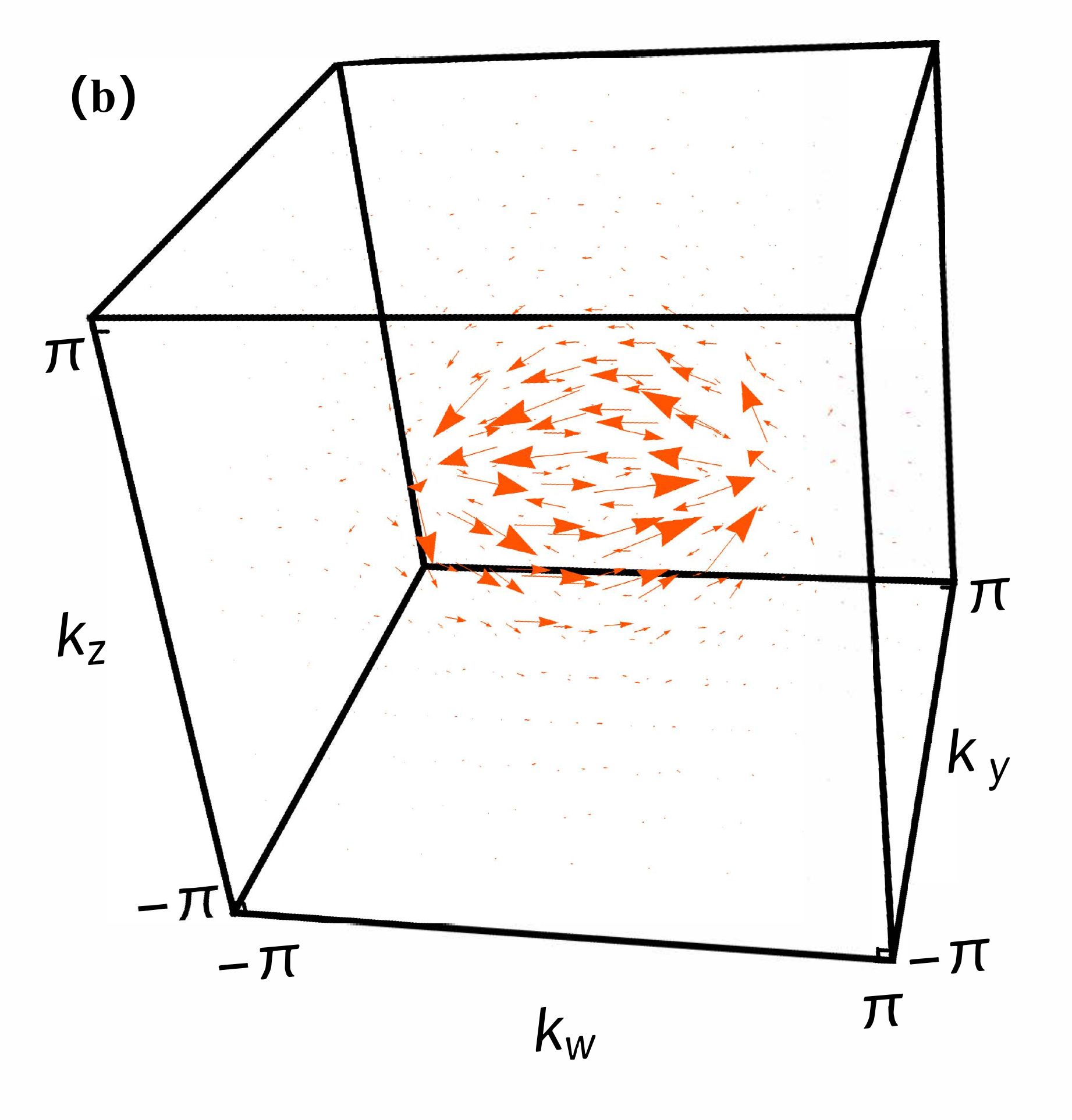}
	\caption{(a) The relative MR with $ n=1 $ and $ h=3 $, $k_x=0$. The magnetic field applied along $ \hat{z} $ direction, the lattice spacing $a=382nm$, $t / \hbar = 2 \pi \times 40 \mathrm{Hz}$, and $E_F=-0.1t$. (b) The vector distribution of the Berry curvature $\mathbf{\Omega}_{\mathbf{k}}$ for $\mathcal{H}_1$ with $k_x=0$.}
	\label{fig5}
\end{figure}
In Fig.~\ref{fig5}, for $ n=1$, $k_x=0$, and $ h=3$, we plot the relative MR of the longitudinal resistance against the magnetic field $ B_z $, which is defined as ~\cite{Du}.
\begin{equation}\label{NMR}
M R_{z}\left(B_{z}\right)=\frac{1 / \sigma_{z z}\left(B_{z}\right)-1 / \sigma_{z z}(0)}{1 / \sigma_{z z}(0)},
\end{equation}
and the results are plotted in Fig.~\ref{fig5}(a), there is a typical $ -B^2 $-dependence of the MR, which signifies the negative MR effect along $\hat{z}$. The distribution of Berry curvature in the momentum space is plotted in Fig.~\ref{fig5}(b), by which we find that the conventional topological charges for both Dirac points vanish. However, the negative MR can still be realized. Intuitively, it is because the coefficient of $ B^2 $ in Eq.~\eqref{E23} is always positive, no matter the integral of $ \mathbf{\Omega}_{\mathbf{k}} $ on the sphere enclosing the Dirac point in 3D is zero. The typical experimental parameters of ultracold atoms in the optical lattice have been used in our calculation, while the magnetic field can be realized by artificial gauge field for the experimental setup. Since the flat zero-energy band doesn't contribute to the conductance because of the vanishing velocity of the wave packet, we don't take the flat band into consideration.


\section{implementation scheme}
\subsection{Realization with optical lattices}
In this subsection, we propose a scheme to realize the 4D Hamiltonian in Eq.~\eqref{tb-H} for $ n=1 $  using ultracold atoms~\cite{Mapping,Pseudos,DWZhang2016,SLZhu2007}. The simulation of this 4D system is achieved by parameterizing the momentum along $ \hat{w} $ on the 3D optical lattice.
For $ n=1 $, we can use noninteracting fermionic atoms in a cubic optical lattice and choose three atomic internal states in the ground state manifold to encode the three spin states $|s\rangle$ $(s=\uparrow,0,\downarrow)$, where the cubic lattice can be formed with three orthogonal sets of counter propagating laser beams with the same wave vector magnitude and the orthogonal polarizations. The tight-binding Hamiltonian of this cold atom system with spin-dependent hopping is written as
\begin{equation}\begin{array}{l}
\hat{H}=t \sum_{\textbf{r}}\left[\hat{H}_{\textbf{rx}}+\hat{H}_{\textbf{ry}}+\hat{H}_{\textbf{rz}}+\hat{H}^{\prime}\right] \\
\hat{H}_{\textbf{rx}}=-i \hat{a}_{\textbf{r-x}, 0}^{+}\left(\hat{a}_{\textbf{r}, \uparrow}+\hat{a}_{\textbf{r}, \downarrow}\right)+i \hat{a}_{\textbf{r+x}, 0}^{+}\left(\hat{a}_{\textbf{r}, \uparrow}-\hat{a}_{\textbf{r}, \downarrow}\right)+H.c. \\
\hat{H}_{\textbf{ry}}=\hat{a}_{\textbf{r-y}, 0}^{+}\left(\hat{a}_{\textbf{r}, \uparrow}-i \hat{a}_{\textbf{r}, \downarrow}\right)-\hat{a}_{\textbf{r+y}, 0}^{+}\left(\hat{a}_{\textbf{r}, \uparrow}+i \hat{a}_{\textbf{r}, \downarrow}\right)+H.c.  \\
\hat{H}_{\textbf{rz}}=-2i\hat{a}_{\textbf{r-z}, 0}^{+}\hat{a}_{\textbf{r}, \downarrow}+H.c. \\
\hat{H}^{\prime}=2i\gamma \hat{a}_{\textbf{r}, 0}^{+}\hat{a}_{\textbf{r}, \downarrow}+H.c.,
\end{array}\end{equation}
where $\hat{H}_{\textbf{rx}}$, $\hat{H}_{\textbf{ry}}$ and  $\hat{H}_{\textbf{rz}}$ represent the hoppings along the $x$, $y$ and $z$ axis, respectively, with the tunneling amplitude $t$ and on-site flipping amplitude $\gamma$. $\hat{a}_{\textbf{r},s}$ and ${\hat{a}^+}_{\textbf{r},s}$ stand for the annihilation and creation operators on lattice site $\textbf{r}$ for the spin state $|s\rangle$ .
In the tight-binding model, the spin-dependent atomic hopping between two nearest neighborhood sites can be realized by Raman coupling between their three spin states~\cite{Zhu}.  Here define the on-site modulated parameter $\gamma=h-\cos\theta$, $h$ is a constant and $\theta$ is a cyclical parameter that vary from $\theta=-\pi$ to $\theta=\pi$.

The generalized 3D tight-binding model on a simple cubic lattice Hamiltonian
\begin{equation}
{\color{red}\hat{H}=\sum_{\mathbf{\tilde{k}}, s s^{\prime}} \hat{a}_{\tilde{\mathbf{k}}, s}^{+}[\mathcal{H}_1(\tilde{\mathbf{k}},\theta)]_{s s'} \hat{a}_{\tilde{\mathbf{k}}, s'},}
\end{equation}
where $\mathbf{\tilde{k}}=(k_x,k_y,k_z)$. By treating the parameter $\theta$ as the pseudo-momentum $k_w$, $\mathcal{H}_1(\mathbf{k})$ is Bloch Hamiltonian as in Eq.~\eqref{tb-H} in a 4D parameter space $\mathbf{k}=(\tilde{\mathbf{k}},\theta)$.

\subsection{Detecting the quantum metric tensor and Berry curvature}

We now turn to address an experimental method to detect the tensor monopole and measure negative magnetoresistance in our system  which is related to measure the quantum metric and Berry curvature in an optical lattice.  Quantum metric tensor is the real part of the quantum geometric tensor, whose imaginary part is just Berry curvature~\cite{Mehta} and has been directly measured in some engineered systems~\cite{Nascimbene,photonics,Rem,Duca,79,80}. In our paper,
the basic experimental procedure is preparing the system in a given Bloch state and introducing an external drive by shaking the lattice~\cite{Tran,periodically,Mei,Reitter,Flaschner,Aidelsburger,FMei2014,GLiu2010,SLZhu2006}. Then the quantum metric and Berry curvature can be measured by establishing its relationship with integrated excitation rate, which is a measurable quantity in experiments~\cite{and,Wilkens,geometrical,Dichroic,Tran,Grushin,quantum}.

In order to measure the quantum metric tensor related to Eq.~\eqref{tb-H} with $ n=1 $, the system is first prepared in the state of $e^{i\mathbf{k}^0 \cdot \mathbf{r}}\left|u_{-}(\mathbf{k}^{0})\right\rangle$. Shaking the lattice along the $x$ direction results in a circular time-periodic perturbation given by
\begin{equation}
\hat{H}_{\hat{x}}(t)=\mathcal{H}_1+2 E\hat{x} \cos (\omega t),
\end{equation}
where $E$ is drive amplitude, $\omega$ is the frequency of shaking driving interband transitions~\cite{and,Tran}. After introducing this external drive, the excitation rate is given by
\begin{equation}
\begin{aligned}
\Gamma_{\hat{x}}(\omega)&= \sum_{n=0,+}\left|\left\langle u_{n}(\mathbf{k})\left|e^{-i \mathbf{k} \cdot \mathbf{r}} \hat{x} e^{i \mathbf{k}^{0} \cdot \mathbf{r}}\right| u_{-}\left(\mathbf{k}^{0}\right)\right\rangle\right|^{2} \\
& \times \frac{2 \pi E^{2}}{\hbar}\delta\left(\hbar \omega^{\prime}\right),
\end{aligned}
\end{equation}
which represents the probability of observing the system in other eigenstates per unit of time. Here $\hbar \omega'=E_n-E_{-}-\hbar \omega$.  By integrating the rate over $\omega$, we obtain
\begin{equation}\Gamma_{\hat{x}}^{\mathrm{int}} =\frac{2 \pi E^{2}}{\hbar^{2}} \sum_{n=0,+}\left|\left\langle u_{n}\left(\mathbf{k}^{0}\right) \mid \partial_{k_{x}} u_{-}\left(\mathbf{k}^{0}\right)\right\rangle\right|^{2}.\end{equation}
The relation between the integrated excitation rate and the quantum metric tensor is now given by
\begin{equation}
\label{30}
\Gamma_{\hat{x}}^{i n t}=\int \Gamma_{\hat{x}}(w) d \omega=\frac{2 \pi E^{2}}{\hbar^{2}} g_{x x}(\mathbf{k}^0),
\end{equation}
where
the diagonal component of quantum metric tensor $g_{xx}=\sum_{n=0,+}\left|\braket{u_n}{\partial_{k_x}u_{-}}\right|^2$.
The relation in Eq.~\eqref{30} provides an experimentally feasible approach to measure the quantum metric tensor.
In concrete practice, one can change the frequency to get the integrated excitation rate as~\cite{Tran,Aidelsburger,Werner}
\begin{equation}
\Gamma_{\hat{x}}^{i n t}=\sum_{i} \Gamma_{\hat{x}}\left(\omega_{i}\right) \Delta \omega.
\end{equation}
To obtain the off-diagonal components of quantum metric tensor, we can measure the excitation rate by applying the shaking along the different directions. Taking $ g_{yz} $ as an example, the shaking can be applied along the directions $\hat{y} \pm \hat{z}$~\cite{and} and then the total Hamiltonian can be written as
\begin{equation}
\label{E32}
\hat{H}_{\hat{y} \pm \hat{z}}(t)=\mathcal{H}_1+2 E(\hat{y} \pm \hat{z}) \cos (\omega t),
\end{equation}
from which we can obtain the excitation rates $ \Gamma_{\hat{y}\pm \hat{z}}^{\text{int}} $ and the difference of those two excitation rates is related to the off-diagonal quantum matric tensor as
\begin{equation}
\Gamma_{\hat{y}+\hat{z}}^{\mathrm{int}}-\Gamma_{\hat{y}-\hat{z}}^{\mathrm{int}}=\frac{8 \pi E^{2}}{\hbar^{2}} g_{y z}.
\end{equation}
By shaking the optical lattice, the topological properties of tensor monopoles can be derived through the measurement of quantum metric tensor. For our 4D system, the generalized Berry curvature $\mathcal{H}_{\mu \nu \lambda}=4\epsilon_{\mu \nu \lambda} \sqrt{\det g_{\mu\nu}}$ can be obtained through extracting the quantum metric tensor, with which the topological charge can be obtained consequently.

Similarly, one can  extract the non-zero component of Berry curvature by simply changing the time-modulation as a circular time-periodic perturbation in Eq.~\eqref{E32}, e.g.,
\begin{equation}
\hat{H}_{\pm}(t)=\mathcal{H}_1+2 E[\cos(\omega t)\hat{y} \pm \sin(\omega t) \hat{z}].
\end{equation}
The corresponding excitation rate derived in Ref.~\cite {Tran} takes the following form,
\begin{equation}
\begin{aligned}
\Gamma_{\pm}(\mathbf{k}; \omega)\hspace{-1mm}&=\hspace{-1mm}\frac{2 \pi}{\hbar} \hspace{-1.8mm} \sum_{n=0,+}\left
|\mathcal{V}_{n -}^{\pm}(\mathbf{k})\right|^{2} \delta^{(t)}\left(E_{n}(\mathbf{k})-E_{-}(\mathbf{k})-\hbar \omega\right)  \\
\left|\mathcal{V}_{n -}^{\pm}(\mathbf{k})\right|^{2}\hspace{-1mm}&=\hspace{-1mm}(\frac{E}{\hbar \omega})^{2}\left|\left\langle u_{n}(\mathbf{k})\left|\frac{1}{i} \frac{\partial \mathcal{H}_{1}}{\partial k_{y}} \mp \frac{\partial \mathcal{H}_{1}}{\partial k_{z}}\right| u_{-}(\mathbf{k})\right\rangle\right|^{2},
\end{aligned}
\end{equation}
and  $\delta^{(t)}(\varepsilon)=(2\hbar/\pi t)\sin ^{2}(\varepsilon t / 2 \hbar) / \varepsilon^{2} \rightarrow \delta(\varepsilon)$ in the long-time limit. Integrating the excitation rate $\Gamma_{\pm}(\mathbf{k}, \omega)$ overall drive frequencies $\omega \geq \Delta_{\mathrm{gap}}/\hbar$($\Delta_{\mathrm{gap}}$ denotes the band gap) and consider the  difference between these integrated rates, which reads
\begin{equation}
\begin{aligned}
\Delta \Gamma^{\mathrm{int}}(\mathbf{k})\!&=\!4 \pi(\frac{E}{\hbar})^{2} \operatorname{Im}\!\!\!\! \sum_{n=0,+} \!\! \! \frac{\left\langle u_{-}\!\left|\partial_{k_{y}} \mathcal{H}_{1}\right| \! u_{n}\right\rangle \!\! \left\langle u_{n} \! \left|\partial_{k_{z}} \mathcal{H}_{1}\right| \! u_{-}\right\rangle}{\left(E_{-}-E_{n}\right)^{2}}  \\
&=-2 \pi(\frac{E}{\hbar})^{2} \Omega_{k_y,k_z},
\end{aligned}
\end{equation}
where $\Omega_{k_y,k_z}$ is one of the components of Berry curvature. This equation gives a feasible approach to measure $\Omega_{k_y,k_z}$, and other components can also be extracted in similar method.
With the result of the Berry curvature,  one could numerically obtain the longitudinal conductivity $\sigma_{zz}$ from  Eq.~\eqref{E23}.

\section{Conclusion}
In summary, we have proposed two minimal Hamiltonians, which host tensor monopoles with topological charges equal to $ n $, and discuss the topological properties of them.  The topological properties and the phase transitions of the tensor monopoles with $ n=1,2 $ have been considered. By increasing $ h $ from zero, the tensor monopoles can be annihilated in pairs of opposite topological charges to open a gap. As $ h>4 $, all tensor monopoles disappear and the system becomes a trivial insulator. The semiclassical Boltzmann equation has been used to calculate the longitudinal conductivity with the  magnetic field, a $-B^2$-dependence of MR is obtained as a result of the Weyl semimetal with a hyperplane cutting through the two tensor monopoles.
An experimental scheme of the topological charge $ 1 $ has been proposed.
We suggest to simulate the 4D Hamiltonian of tensor monopole by the 3D optical lattice with a parametrized pseudo-momentum along the fourth dimension. The relation between the total excitation rate and the quantum metric tensor facilitates us to measure the quantum metric tensor by shaking the optical lattice.

\begin{acknowledgments}
	We thank  G. Palumbo, N. Goldman, D. W. Zhang, and S. L. Zhu for helpful discussions. This work was supported by the National Key Research and Development Program of China (Grant No. 2016YFA0301800), the National Nature Science Foundation of China (Grant No. 11704180, 11474153), the Key Project of Science and  Technology of Guangzhou (Grant No. 201804020055) and Key R\&D Program of Guangdong province (Grant No. 2019B030330001).\end{acknowledgments}

\begin{appendix}
\section{Calculation of the surface state spectrum}
Expand the Hamiltonian $\mathcal{H}_2$ around $(k_x,k_y,k_z)=(0,0,0)$, and consider the open boundary condition along $z$ direction, the Hamiltonian can be rewritten as
\begin{equation}
\mathcal{H}=\left(\begin{array}{ccc}{0} & {\hat{d}_{1}-i\hat{d}_{2}} & {0} \\ {\hat{d}_{1}+i\hat{d}_{2}} & {0} & {\hat{d}_{3}+i A_{000}} \\ {0} & {\hat{d}_{3}-i A_{000}} & {0}\end{array}\right),
\end{equation}
where
\begin{equation}
\begin{split}
\hat{d}_{1}&=2t(k_x^2-k_y^2),\\
\hat{d}_{2}&=4tk_x k_y,\\
\hat{d}_{3}&=2tk_z=-2i t\partial_{z},\\
A_{000}&=2t(h-3-\cos k_w).
\end{split}
\label{hs}
\end{equation}
Define $A_{000}\equiv \hat{d}_{4}$, and we regard $A_{000}$ as a domain wall  configuration along the $z$-direction, which we choose to parametrize as
\begin{equation}
A_{000}\left(z\right)=\bar{A}_{000}\left[\Theta\left(z\right)-\Theta\left(-z\right)\right],
\end{equation}
Here $\bar{A}_{000}=-A_{000}$, and $\Theta$ is the  Heaviside function with
\begin{equation}
\Theta(z)=\left\{\begin{array}{ll}
{1} & {, z>0} \\
{\frac{1}{2}} & {, z=0} \\
{0} & {, z<0}.
\end{array}\right.
\end{equation}
Since $k_x$ and $k_y$ are good quantum numbers, we can use their eigenvalues to replace the momentum operators, and solve eigen-equation
\begin{equation}
\mathcal{H} \psi=\varepsilon_{k} \psi,
\end{equation}
with $\psi=e^{ik_x x+ik_y y} \phi(z)$. The components of the spinor wavefunction $\phi\left(z\right)=\left(f\left(z\right), g\left(z\right), h\left(z\right)\right)^{\top}$. Combining above equations derive
\begin{equation}
\begin{aligned}
&f(z)=\frac{1}{\varepsilon_{k}}\left(\hat{d}_{1}-i \hat{d}_{2}\right) g( z),\\
&h(z)=\frac{1}{\varepsilon_{k}}\left(-2 i \partial_{z}-i A_{000}\right) g(z),
\end{aligned}
\end{equation}
and
\begin{equation}
\left[-4 \partial_{z}^{2}+A_{000}^2-4  \bar{A}_{000}\delta(z)\right] g(z)=\left[\varepsilon_{k}^{2}-\left(\hat{d}_{1}^{2}+\hat{d}_{2}^{2}\right)\right]g(z),
\end{equation}
at $z \neq 0$. The solution of Eq. A(7) is
\begin{equation}
h\left(z\right)=h_{0} e^{-\left|z\right| / \lambda},
\end{equation}
where $h_0$ is a  normalization constant and $\lambda^{-1}:=\sqrt{\bar{A}_{000}^{2}+(\hat{d}_1^2+\hat{d}_2^2)-\varepsilon_{\kappa}^{2}}>0$, the discontinuity of delta function  at $z=0$ imposes the condition $\lambda^{-1}={\bar{A}_{000}}$. Therefore, the surface states dispersions are given by
\begin{equation}
\varepsilon_{\pm,k}=\pm \sqrt{(\hat{d}_1^2+\hat{d}_2^2)}=\pm v(k_x^2+k_y^2),
\end{equation}
where $v=2t$ is the effective Fermi velocity. For $\bar{A}_{000}>0$, we only consider $h=3$ in the main text, so we derive $k_w \in (-\pi/2,\pi/2)$.  The surface state spectrum of $\mathcal{H}_1$ can be derived in the same method.

\end{appendix}

\bibliography{m1}

\end{document}